\RequirePackage{lineno}
\documentclass[aps,prd,twocolumn,showpacs,superscriptaddress]{revtex4}  
\usepackage{graphicx}  
\usepackage{dcolumn}   
\usepackage{bm}        
\usepackage{amssymb}   
\usepackage{amsmath}

\begin{document}

\hspace{5.2in} \mbox{FERMILAB-PUB-11/206-E}

\title{Measurement of the production fraction times branching fraction $\boldsymbol{ f(b\to\Lambda_{b})\cdot \mathcal{B}(\Lambda_{b}\to J/\psi \Lambda)}$}

\affiliation{Universidad de Buenos Aires, Buenos Aires, Argentina}
\affiliation{LAFEX, Centro Brasileiro de Pesquisas F{\'\i}sicas, Rio de Janeiro, Brazil}
\affiliation{Universidade do Estado do Rio de Janeiro, Rio de Janeiro, Brazil}
\affiliation{Universidade Federal do ABC, Santo Andr\'e, Brazil}
\affiliation{Instituto de F\'{\i}sica Te\'orica, Universidade Estadual Paulista, S\~ao Paulo, Brazil}
\affiliation{Simon Fraser University, Vancouver, British Columbia, and York University, Toronto, Ontario, Canada}
\affiliation{University of Science and Technology of China, Hefei, People's Republic of China}
\affiliation{Universidad de los Andes, Bogot\'{a}, Colombia}
\affiliation{Charles University, Faculty of Mathematics and Physics, Center for Particle Physics, Prague, Czech Republic}
\affiliation{Czech Technical University in Prague, Prague, Czech Republic}
\affiliation{Center for Particle Physics, Institute of Physics, Academy of Sciences of the Czech Republic, Prague, Czech Republic}
\affiliation{Universidad San Francisco de Quito, Quito, Ecuador}
\affiliation{LPC, Universit\'e Blaise Pascal, CNRS/IN2P3, Clermont, France}
\affiliation{LPSC, Universit\'e Joseph Fourier Grenoble 1, CNRS/IN2P3, Institut National Polytechnique de Grenoble, Grenoble, France}
\affiliation{CPPM, Aix-Marseille Universit\'e, CNRS/IN2P3, Marseille, France}
\affiliation{LAL, Universit\'e Paris-Sud, CNRS/IN2P3, Orsay, France}
\affiliation{LPNHE, Universit\'es Paris VI and VII, CNRS/IN2P3, Paris, France}
\affiliation{CEA, Irfu, SPP, Saclay, France}
\affiliation{IPHC, Universit\'e de Strasbourg, CNRS/IN2P3, Strasbourg, France}
\affiliation{IPNL, Universit\'e Lyon 1, CNRS/IN2P3, Villeurbanne, France and Universit\'e de Lyon, Lyon, France}
\affiliation{III. Physikalisches Institut A, RWTH Aachen University, Aachen, Germany}
\affiliation{Physikalisches Institut, Universit{\"a}t Freiburg, Freiburg, Germany}
\affiliation{II. Physikalisches Institut, Georg-August-Universit{\"a}t G\"ottingen, G\"ottingen, Germany}
\affiliation{Institut f{\"u}r Physik, Universit{\"a}t Mainz, Mainz, Germany}
\affiliation{Ludwig-Maximilians-Universit{\"a}t M{\"u}nchen, M{\"u}nchen, Germany}
\affiliation{Fachbereich Physik, Bergische Universit{\"a}t Wuppertal, Wuppertal, Germany}
\affiliation{Panjab University, Chandigarh, India}
\affiliation{Delhi University, Delhi, India}
\affiliation{Tata Institute of Fundamental Research, Mumbai, India}
\affiliation{University College Dublin, Dublin, Ireland}
\affiliation{Korea Detector Laboratory, Korea University, Seoul, Korea}
\affiliation{CINVESTAV, Mexico City, Mexico}
\affiliation{FOM-Institute NIKHEF and University of Amsterdam/NIKHEF, Amsterdam, The Netherlands}
\affiliation{Radboud University Nijmegen/NIKHEF, Nijmegen, The Netherlands}
\affiliation{Joint Institute for Nuclear Research, Dubna, Russia}
\affiliation{Institute for Theoretical and Experimental Physics, Moscow, Russia}
\affiliation{Moscow State University, Moscow, Russia}
\affiliation{Institute for High Energy Physics, Protvino, Russia}
\affiliation{Petersburg Nuclear Physics Institute, St. Petersburg, Russia}
\affiliation{Instituci\'{o} Catalana de Recerca i Estudis Avan\c{c}ats (ICREA) and Institut de F\'{i}sica d'Altes Energies (IFAE), Barcelona, Spain}
\affiliation{Stockholm University, Stockholm and Uppsala University, Uppsala, Sweden}
\affiliation{Lancaster University, Lancaster LA1 4YB, United Kingdom}
\affiliation{Imperial College London, London SW7 2AZ, United Kingdom}
\affiliation{The University of Manchester, Manchester M13 9PL, United Kingdom}
\affiliation{University of Arizona, Tucson, Arizona 85721, USA}
\affiliation{University of California Riverside, Riverside, California 92521, USA}
\affiliation{Florida State University, Tallahassee, Florida 32306, USA}
\affiliation{Fermi National Accelerator Laboratory, Batavia, Illinois 60510, USA}
\affiliation{University of Illinois at Chicago, Chicago, Illinois 60607, USA}
\affiliation{Northern Illinois University, DeKalb, Illinois 60115, USA}
\affiliation{Northwestern University, Evanston, Illinois 60208, USA}
\affiliation{Indiana University, Bloomington, Indiana 47405, USA}
\affiliation{Purdue University Calumet, Hammond, Indiana 46323, USA}
\affiliation{University of Notre Dame, Notre Dame, Indiana 46556, USA}
\affiliation{Iowa State University, Ames, Iowa 50011, USA}
\affiliation{University of Kansas, Lawrence, Kansas 66045, USA}
\affiliation{Kansas State University, Manhattan, Kansas 66506, USA}
\affiliation{Louisiana Tech University, Ruston, Louisiana 71272, USA}
\affiliation{Boston University, Boston, Massachusetts 02215, USA}
\affiliation{Northeastern University, Boston, Massachusetts 02115, USA}
\affiliation{University of Michigan, Ann Arbor, Michigan 48109, USA}
\affiliation{Michigan State University, East Lansing, Michigan 48824, USA}
\affiliation{University of Mississippi, University, Mississippi 38677, USA}
\affiliation{University of Nebraska, Lincoln, Nebraska 68588, USA}
\affiliation{Rutgers University, Piscataway, New Jersey 08855, USA}
\affiliation{Princeton University, Princeton, New Jersey 08544, USA}
\affiliation{State University of New York, Buffalo, New York 14260, USA}
\affiliation{Columbia University, New York, New York 10027, USA}
\affiliation{University of Rochester, Rochester, New York 14627, USA}
\affiliation{State University of New York, Stony Brook, New York 11794, USA}
\affiliation{Brookhaven National Laboratory, Upton, New York 11973, USA}
\affiliation{Langston University, Langston, Oklahoma 73050, USA}
\affiliation{University of Oklahoma, Norman, Oklahoma 73019, USA}
\affiliation{Oklahoma State University, Stillwater, Oklahoma 74078, USA}
\affiliation{Brown University, Providence, Rhode Island 02912, USA}
\affiliation{University of Texas, Arlington, Texas 76019, USA}
\affiliation{Southern Methodist University, Dallas, Texas 75275, USA}
\affiliation{Rice University, Houston, Texas 77005, USA}
\affiliation{University of Virginia, Charlottesville, Virginia 22901, USA}
\affiliation{University of Washington, Seattle, Washington 98195, USA}
\author{V.M.~Abazov} \affiliation{Joint Institute for Nuclear Research, Dubna, Russia}
\author{B.~Abbott} \affiliation{University of Oklahoma, Norman, Oklahoma 73019, USA}
\author{B.S.~Acharya} \affiliation{Tata Institute of Fundamental Research, Mumbai, India}
\author{M.~Adams} \affiliation{University of Illinois at Chicago, Chicago, Illinois 60607, USA}
\author{T.~Adams} \affiliation{Florida State University, Tallahassee, Florida 32306, USA}
\author{G.D.~Alexeev} \affiliation{Joint Institute for Nuclear Research, Dubna, Russia}
\author{G.~Alkhazov} \affiliation{Petersburg Nuclear Physics Institute, St. Petersburg, Russia}
\author{A.~Alton$^{a}$} \affiliation{University of Michigan, Ann Arbor, Michigan 48109, USA}
\author{G.~Alverson} \affiliation{Northeastern University, Boston, Massachusetts 02115, USA}
\author{G.A.~Alves} \affiliation{LAFEX, Centro Brasileiro de Pesquisas F{\'\i}sicas, Rio de Janeiro, Brazil}
% ...
%
%
\collaboration{The D0 Collaboration\footnote{with visitors from
$^{a}$Augustana College, Sioux Falls, SD, USA,
$^{b}$The University of Liverpool, Liverpool, UK,
$^{c}$SLAC, Menlo Park, CA, USA,
$^{d}$University College London, London, UK,
$^{e}$Centro de Investigacion en Computacion - IPN, Mexico City, Mexico,
$^{f}$ECFM, Universidad Autonoma de Sinaloa, Culiac\'an, Mexico,
and 
$^{g}$Universit{\"a}t Bern, Bern, Switzerland.
}} \noaffiliation
\vskip 0.25cm
\date{May 3, 2011}

\begin{abstract}

The $\Lambda_{b}(udb)$ baryon is observed in the decay $\Lambda_{b}\to J/\psi \Lambda$ using 
6.1 fb$^{-1}$ of $p\bar{p}$ collisions collected with the 
D0 detector at $\sqrt{s}=1.96$ TeV. The production fraction multiplied by the branching fraction for this
decay relative to that for the decay    $B^{0}\to J/\psi K^{0}_{s}$  is measured to be 
$0.345 \pm 0.034 \mbox{ (stat.)} \pm 0.033\mbox{ (syst.)} \pm 0.003\mbox{ (PDG)}$. 
Using the  world average value of $f(b\to B^{0})\cdot\mathcal{B}(B^{0}\to J/\psi K^{0}_{s})=(1.74\pm 0.08)\times 10^{-5}$, we obtain 
 $f(b\to\Lambda_{b})\cdot \mathcal{B} \left( \Lambda_{b} \rightarrow  J/\psi  \Lambda \right) = (6.01\pm 0.60\mbox{ (stat.)}\pm0.58\mbox{ (syst.)} \pm  0.28 \mbox{ (PDG)})\times 10^{-5}$.
This measurement represents an improvement in precision by about a factor of three with respect to the current  world average.

\end{abstract}

\maketitle

The study of $b$ hadron decays, in particular  $b\to s$ decays, 
offers good opportunities to search for physics beyond the standard model (BSM).
For this reason, these decays have been the subject of intensive experimental~\cite{ExpW1,ExpW2,ExpW2a,ExpW3,ExpW3a,ExpW4} and theoretical~\cite{TheW1,TheW2,TheW3} work.  
Studies of $b$ baryons at the Fermilab Tevatron Collider and the CERN Large Hadron Collider are a natural extension of these studies which have been mostly performed on $B$ mesons~\cite{Ext,Ext1,Ext2,Ext4}.
The experimental knowledge of $b$  baryons is currently  limited~\cite{PDG}. 
For   the $\Lambda_{b}(udb)$, 
the lightest  $b$ baryon, only a few decay channels have been studied, and the uncertainties on its branching fractions  are large
$\sim$($30$--$60$)\%. 
For higher mass $b$ baryon states, even less information is available.
Due to its relative abundance, the $\Lambda_b$ baryon has
been used to investigate  production and
decay properties of heavier $b$ baryons, to search for possible
polarization effects ~\cite{Pol1}, for violation of
discrete symmetries in the decay (CP~\cite{CPLb}
and T~\cite{TLb} violation), and to search for BSM effects~\cite{NewPhysLb}.
There  are  several models 
(PQCD~\cite{PQCD}, relativistic and non-relativistic quark models based on factorization aproximations~\cite{Fact1,Fact2,Fact3,Fact4,Fact5,Fact6} 
are  examples)
to describe $b$ baryon decays such as $\Lambda_b \rightarrow J/\psi \Lambda$.
Increasingly precise measurements of $f(b\to\Lambda_{b})\cdot \mathcal{B}(\Lambda_{b}\to J/\psi \Lambda)$ (where $f(b\to\Lambda_{b})$ 
is the fraction of $b$ quarks which hadronize to $\Lambda_{b}$ baryons)
will allow  better tests of these  models.
Moreover, these measurements could help in the study of 
$b\to s$ decays such as $\Lambda_{b}\to \mu^{+} \mu^{-} \Lambda$ ~\cite{Lbrare1,Lbrare2},
 which are topologically similar to 
$\Lambda_{b}\to J/\psi \Lambda$, where $J/\psi$ decays to dimuons.

This Letter reports an improved  measurement with respect to the previous Tevatron result~\cite{CDFBrLbrecent} of the 
production fraction multiplied by the branching fraction of the $\Lambda_{b}\to J/\psi \Lambda$ decay relative to that of the decay 
$B^{0}\to J/\psi K^{0}_{s}$.  From this measurement we can obtain 
$ f(b\to\Lambda_{b})\cdot \mathcal{B}(\Lambda_{b}\to J/\psi \Lambda)$ with
significantly improved precision  compared to the current world average~\cite{PDG}.
The $J/\psi$, $\Lambda$,  and  $K^{0}_{s}$ are  reconstructed in the $\mu^{+}\mu^{-}$, $p\pi^{-}$, and $\pi^{+}\pi^{-}$ modes, respectively.
Throughout this Letter, the appearance of a specific charge state also implies its charge conjugate.
The study is performed using 6.1 fb$^{-1}$ of $p\bar{p}$ collisions collected with the 
D0 detector between 2002--2009 at $\sqrt{s}=1.96$ TeV at the Fermilab Tevatron Collider.

A detailed description of the D0 detector can be found in~\cite{d0det}.
The components most relevant to this analysis are the central tracking system and the muon spectrometer. The central 
tracking system consists of a silicon microstrip tracker (SMT) and a central fiber tracker (CFT) that are surrounded 
by a 2~T superconducting solenoid. The SMT is optimized for tracking and vertexing for the pseudorapidity region
 $|\eta|<3.0$ (where $\eta=-\ln[\tan(\theta/2)]$ and $\theta$ is the polar angle), while the CFT has coverage for $|\eta|<2.0$.
 Liquid-argon and uranium calorimeters in a central and two end-cap cryostats cover the pseudorapidity region $|\eta|<4.2$.
 The muon spectrometer is located outside the calorimeter and covers   $|\eta|<2.0$. It comprises a 
layer of drift tubes and scintillator trigger counters in front of 1.8~T iron toroids followed by two similar layers after the toroids.

We closely follow the
data selection for $J/\psi\to\mu^{+}\mu^{-}$, $\Lambda\to p\pi^{-}$
and $K^{0}_{s}\to \pi^{+}\pi^{-}$   used in the measurement~\cite{Lifetime}
of the ratio of the lifetimes, $\tau(\Lambda_{b})/\tau(B^{0})$,
that  used  the same decay products of the $\Lambda_b$
and $B^0$. Events satisfying muon or dimuon triggers are used.
At least one $p\bar{p}$ interaction vertex must  be identified in each event, determined by  minimizing a  $\chi^2$ function 
that depends on all reconstructed tracks in the event and a term that represents the average beam position constraint.
We begin by 
searching   for $J/\psi\to\mu^+\mu^-$ decays
reconstructed from two oppositely charged muons that have a common 
vertex with a $\chi^2$ probability greater than 1\%.
Muons are identified by matching tracks reconstructed in the central tracking system with  
track segments in the muon spectrometer. 
The requirements  of transverse momentum $p_T>2.0$~GeV/$c$ and $|\eta|<2.0$ are imposed on these matched tracks, and each of them
must be associated to at least two hits in the SMT and two hits in the CFT.  In addition, at least one muon track must have segments in the muon system both inside and outside the toroid.
The dimuon transverse momentum $p_{T}(\mu^{+}\mu^{-})$ is required to be greater 
than $3.0$ GeV/$c$, and its invariant mass $M_{\mu^{+}\mu^{-}}$ must be in the range $2.8-3.35$~GeV/$c^2$. 
In these dimuon events we search for $\Lambda\to p\pi^-$  and 
$K^{0}_{s}\to\pi^{+}\pi^{-}$ candidates formed from two oppositely charged tracks with a 
common vertex with a $\chi^2$ probability greater than 1\% and invariant mass between 
1.102 $ < M_{\Lambda} < $ 1.130~GeV/$c^2$   and  0.466 $ < M_{ K^{0}_{S} } < $  0.530 GeV/$c^2$.
To reduce the contribution from fake vertices reconstructed from random track crossings,
the two tracks are required to have at most
two hits associated to them in the tracking detectors
located between the reconstructed $p\bar{p}$ interaction vertex and the
common two-track vertex.
The impact parameter significance (the impact parameter 
with respect to the $p\bar{p}$ vertex
divided by its uncertainty)
for the tracks forming $\Lambda$ or $K^0_S$ candidates must 
exceed $3$ for both tracks and $4$ for at least one of them. 
To reconstruct $\Lambda$ candidates, the track with the higher $p_T$
is assumed to be a proton. 
Monte Carlo studies show that this is always the correct assignment, given the track $p_{T}$ detection threshold.
To suppress contamination from cascade decays of more massive baryons such as
$\Sigma^{0}\to\Lambda\gamma$ and $\Xi^{0}\to\Lambda\pi^{0}$, we require the
cosine of the angle between the $p_T$ of the $\Lambda$
and the vector from the $J/\psi$  vertex to the $\Lambda$ decay vertex
in the plane perpendicular to the beam direction
to be larger than 0.999. For $\Lambda$ candidates coming from
$\Lambda_{b}$ decays, the cosine of this angle is typically greater that 0.9999.

The $\Lambda_b$  ($B^0$) is reconstructed by performing a constrained 
fit to a common vertex for  the $\Lambda$ ($K^0_S$) candidate and the two muon
tracks, with the muons constrained to the nominal $J/\psi$ mass of 3.097~GeV/$c^2$~\cite{PDG}. 
The $p_{T}$ of the $\Lambda_{b}$ or $B^0$ candidate is required to be greater than 5 GeV/$c$. 
The invariant mass of the $J/\psi$ and the two additional tracks is required to be
within the range 5.0--6.2~GeV/$c^2$ for $\Lambda_b$ candidates and 
within 4.8--5.8~GeV/$c^2$ for $B^0$ candidates.

To determine the final selection criteria, we maximize $N_{S}/\sqrt{N_{S}+N_{B}}$, where $N_{S}$ is  
the number of signal  ($\Lambda_{b}$ or $B^{0}$) candidates determined by Monte Carlo  and $N_{B}$  is the number of  background candidates estimated
by using data events in the sidebands of the expected signal. For the Monte Carlo, 
we use {\sc pythia}~\cite{Pythia} and
{\sc evtgen}~\cite{EvtGen} for the production and decay of the simulated particles, respectively, and {\sc geant3}~\cite{Geant} 
to simulate detector effects.
As a result of this optimization, for the  $\Lambda$ ($K^0_S$) we require
the transverse decay length to be greater than 0.8 (0.4) cm, the  $p_T$ to be greater
than 1.6 (1.0) GeV/$c$ and 
the significance of its transverse proper decay length 
(transverse decay length corrected by the boost in the transverse plane)
to be greater than 4.0 (9.0). For the $\Lambda_{b}$ ($B^{0}$) candidate,
the significance of the proper decay length is required to be greater than 2.0 (3.0). 
In addition,  the $\Lambda_{b}$ and  $B^{0}$ vertices must be well reconstructed.

A track pair can be simultaneously identified as both $\Lambda$  and $K^0_S$ due to  different mass assignments to the same tracks.
Events containing such track pair ambiguities are removed.
Finally, if more than one candidate is found in the event, the candidate with the best vertex 
$\chi^2$ probability is selected as the $\Lambda_b$ ($B^0$). 
\begin{figure}
\includegraphics[scale=0.43]{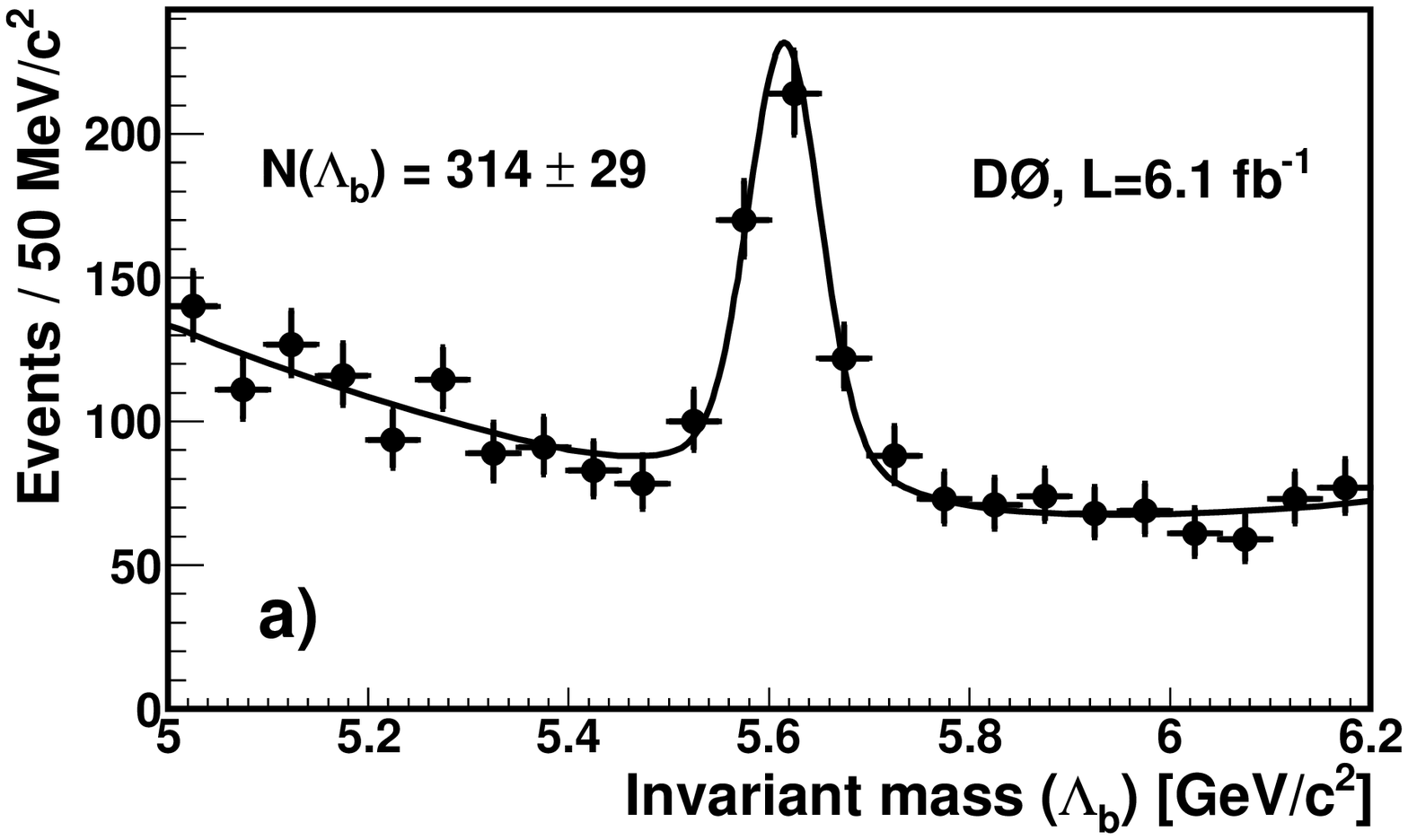}\\
\includegraphics[scale=0.43]{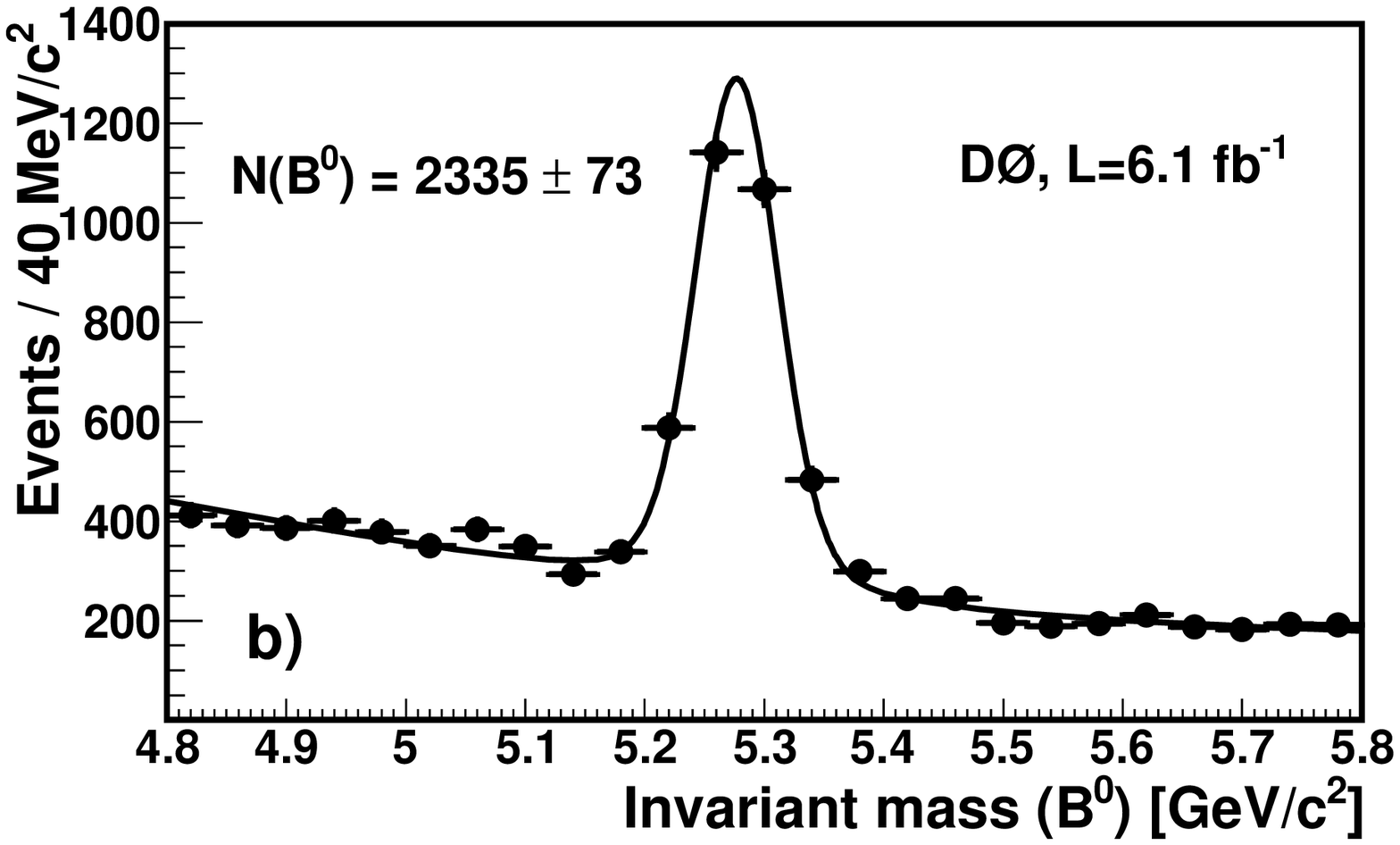}
\caption{\label{fig:mass} Invariant mass distribution in data for  (a)  $\Lambda_{b}\to J/\psi \Lambda$ and (b)
$B^{0}\to J/\psi K^{0}_{s}$  decays. Fit results are superimposed.}
\end{figure}

The invariant mass distributions of the final $\Lambda_{b}$ and $B^{0}$ candidates passing our selection 
criteria are shown in Fig.~\ref{fig:mass}.  To extract the yields of the
observed $\Lambda_{b}$  and $B^{0}$ hadrons, we perform an unbinned likelihood fit to 
each mass distribution assuming a double Gaussian function for signal and a second order polynomial
distribution for background. The 
fits yield
$N_{\Lambda_{b}\to J/\psi\Lambda}=314\pm 29$ events and $N_{B^{0}\to J/\psi K^{0}_{S}}=2335\pm 73$ events.

The relative production fraction times  branching fraction for $\Lambda_{b}\to J/\psi \Lambda$ decays
 to that of $B^{0}\to J/\psi K^{0}_{s}$ decays is given by
\begin{eqnarray}
\sigma_{\text{rel}} &\equiv& \frac{f(b\to\Lambda_{b})\cdot\mathcal{B}(\Lambda_{b}\to J/\psi \Lambda)}{f(b\to B^{0}) \nonumber 
\cdot\mathcal{B}(B^{0}\to J/\psi K^{0}_{s})} \nonumber \\ &=&
\frac{N_{\Lambda_{b}\to J/\psi\Lambda}}{N_{B^{0}\to J/\psi K^{0}_{S}}} 
\cdot \frac{\mathcal{B}(K^{0}_{s}\to\pi^{+}\pi^{-})}{\mathcal{B}(\Lambda\to p\pi^{-})} \cdot \epsilon .
\end{eqnarray}
Here, $\epsilon = \epsilon_{B^{0}\to J/\psi K^{0}_{S}}  / \epsilon_{\Lambda_{b}\to J/\psi\Lambda}$ 
is the relative detection efficiency of $B^{0}\to J/\psi K^{0}_{S}$ to $\Lambda_{b}\to J/\psi\Lambda$ decays.
This relative efficiency is determined from  Monte Carlo (MC) simulation to be $\epsilon=2.37\pm0.05  ~\mbox{(MC stat.)}$. Using
$\mathcal{B}(K^{0}_{s}\to\pi^{+}\pi^{-})=0.6920\pm0.0005$
and $\mathcal{B}(\Lambda\to p\pi^{-})=0.639\pm0.005$~\cite{PDG}, we obtain  
$ \mathcal{B}(K^{0}_{s}\to\pi^{+}\pi^{-}) / \mathcal{B}(\Lambda\to p\pi^{-}) = 1.083 \pm 0.009$. 
With  these inputs and the reconstructed $\Lambda_{b}$ and $B^{0}$ yields, the relative production fraction is found to be 
$\sigma_{\text{rel}}=0.345 \pm 0.034 \mbox{ (stat.)} \pm 0.003 \mbox{ (PDG)}$,
where (PDG) denotes the uncertainty due to the inputs from~\cite{PDG}.

The sources of systematic uncertainty on $\sigma_{\text{rel}}$ are: 
(i)   uncertainties in the determination of the $\Lambda_b$ and
$B^0$ yields,
(ii) the determination of  the relative efficiency $\epsilon$,
(iii)  contamination from $\Lambda_{b}$ in $B^{0}$ and conversely,
 and 
(iv) $\Lambda_{b}$ polarization effects on the relative efficiency $\epsilon$. 
Many other systematic uncertainties common to both $\Lambda_{b}\to J/\psi \Lambda$ and
$B^{0}\to J/\psi K^{0}_{s}$ decays, such as $b$ quark production,
integrated luminosity, trigger and selection efficiencies,
cancel  in the ratio.
The models used for describing signal and
background in data are varied, and the resulting changes
in the $\Lambda_b$ and $B^0$ yields  introduce a
maximum deviation of $\sigma_{\text{rel}}$ from its central
value of 5.5\%, which is included as systematic uncertainty.
The  simulation used to  estimate $\epsilon$  uses a phase space model 
in {\sc evtgen} to decay $\Lambda_{b}$ and $B^{0}$ particles. 
For $B^{0}$ decays we  can also use the  SVS\_CP (Scalar-Vector-Scalar with CP violation) model~\cite{EvtGen}. 
When using this alternative model, we observe a deviation of $2.0$\% in $\sigma_{\text{rel}}$.
Given the similar topologies
 of the $\Lambda_{b}\to J/\psi(\mu^{+}\mu^{-}) \Lambda(p\pi^{-})$ and $B^{0}\to J/\psi(\mu^{+}\mu^{-}) K^{0}_{s}(\pi^{+}\pi^{-})$ 
decays, the  $\Lambda_{b}$ sample may be contaminated  with  $B^{0}$ events that pass the $\Lambda_{b}$ selection, or vice versa. 
We quantify this effect in  simulation and 
find a deviation of $2.3$\% in $\sigma_{\text{rel}}$, which we include as a systematic uncertainty.
Finally, the effect of the unknown polarization and decay parameters of the $\Lambda_{b}$ baryon on the relative efficiency is studied following the formalism of~\cite{Pol1,Pol2}.
The main effect of the polarization is observed through $\Theta$, the emission angle of the $\Lambda$ baryon with 
respect to the polarization direction in the $\Lambda_{b}$ rest frame. This angle follows the distribution
$I(\Theta)\propto 1 + \alpha_{\Lambda_{b}} P_{\Lambda_{b}} \cos(\Theta)$, where $\alpha_{\Lambda_{b}}$  and $P_{\Lambda_{b}}$ are the asymmetry parameter and polarization of the
$\Lambda_{b}$ baryon. We study  the extreme cases $\alpha_{\Lambda_{b}} P_{\Lambda_{b}}=\pm 1$ in simulations. 
The maximum deviation found in $\sigma_{\text{rel}}$ is $7.2\%$, which
is included as a systematic uncertainty due to the unknown $\Lambda_{b}$ polarization. 
All of these systematic uncertainties are combined  assuming  no correlations, giving a total systematic 
uncertainty of $9.6\%$.

\begin{figure}[t]
\includegraphics[scale=0.49]{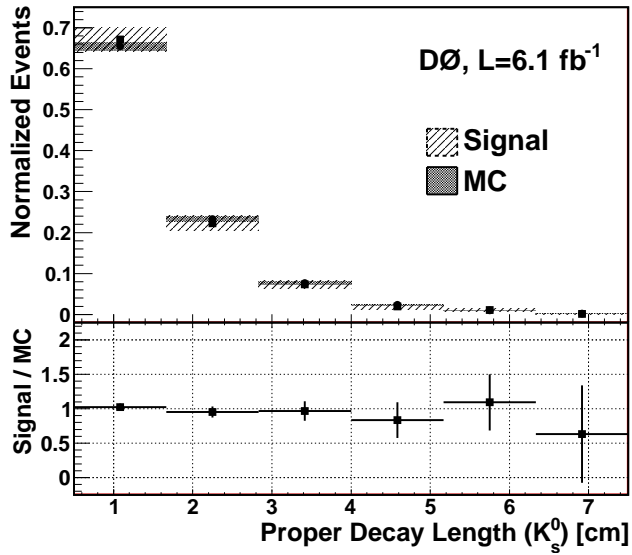}
\caption{\label{fig:K0sct} Proper decay length distributions for $K^{0}_{s}$ candidates in the decay   $B^{0}\to J/\psi K^{0}_{s}$
for background subtracted signal compared to the full Monte Carlo (MC) simulation. The ratio Signal/MC is given 
  in the bottom panel.
}
\end{figure}

We  study the stability
of the measurement by performing cross checks 
on the two main inputs to the computation of $\sigma_{\text{rel}}$: 
the ratio between the numbers of observed $\Lambda_{b}$ and $B^{0}$ candidates extracted from data
and the relative efficiency determined from Monte Carlo. We investigate the possibility that the
number of $\Lambda_b$ and $B^0$ candidates is affected by time
or kinematics dependent changes in the detection and selection
efficiency. We divide the data in subsamples and
determine the value of $\sigma_{\text{rel}}$ in each individual
subsample without observing any significant deviation
from the measurement based on the full sample. We 
split the sample based on different data taking periods,
in different $p_T$, $\eta$ regions,  $\Lambda$ and $K^0_S$ decay lengths   and also investigated
differences between $\Lambda_b$ and $\bar{\Lambda}_b$ rates. To test
for any mismodeling
of the detector efficiency that could affect the determination of $\sigma_{\text{rel}}$,
decay length distributions are compared between data and Monte Carlo, as well as
proper decay length significance, $\chi^{2}$ vertex distributions, and other variables used in the selection.
In all these comparisons, the data and Monte Carlo distributions are found to be in good agreement.  
One such example is Fig.~\ref{fig:K0sct}, which shows the proper decay length~\cite{PDL} distribution of $K^{0}_{S}$ candidates. As a final cross-check, lifetime measurements are performed for the  $\Lambda$ and $K^{0}_{S}$ with results
 in agreement with the world average values~\cite{PDG}.

In summary, using an integrated luminosity of 6.1 fb$^{-1}$  collected with the D0 detector, 
we  measure  the production fraction multiplied by the branching fraction for the decay $\Lambda_{b}\to J/\psi \Lambda$ relative
to that for the decay $B^{0}\to J/\psi K^{0}_{s}$,

\vspace{-0.18cm}
\begin{eqnarray}
\sigma_{\text{rel}} &=& 0.345 \pm 0.034 \mbox{ (stat.)}  \nonumber \\
&& \pm~0.033\mbox{ (syst.)} \pm 0.003\mbox{ (PDG)}.
\end{eqnarray}

Combining the uncertainties  in quadrature, we obtain $\sigma_{\text{rel}} =  0.345 \pm 0.047 $.  Our measurement is the most precise to date and exceeds the precision of the current value reported as the world average,
$0.27\pm 0.13 $~\cite{PDG}.
Using the PDG value   $f(b\to B^{0})\cdot\mathcal{B}(B^{0}\to J/\psi K^{0}_{s})=(1.74\pm 0.08)\times 10^{-4}$ (from~\cite{PDG}),
 we obtain
\begin{linenomath*}
\begin{eqnarray}
\label{Brm}
\lefteqn{f(b\to\Lambda_{b})\cdot\mathcal{B}(\Lambda_{b}\to J/\psi \Lambda) = } \nonumber \\  
&& \left [6.01\pm 0.60\mbox{ (stat.)}  \pm 0.58\mbox{ (syst.)} \right. \nonumber \\
&& \left. {} \pm  0.28 \mbox{ (PDG)} \right ] \times 10^{-5} = (6.01 \pm 0.88)\times 10^{-5},~~~  
\end{eqnarray}
\end{linenomath*}
which can be compared directly to the  world average value of $(4.7\pm 2.3)\times 10^{-5}$~\cite{PDG}.
This result represents a  reduction by a factor of $\sim$3 of the uncertainty
with respect to the  previous  measurement~\cite{CDFBrLbrecent}.

%
We thank the staffs at Fermilab and collaborating institutions,
and acknowledge support from the
DOE and NSF (USA);
CEA and CNRS/IN2P3 (France);
FASI, Rosatom and RFBR (Russia);
CNPq, FAPERJ, FAPESP and FUNDUNESP (Brazil);
DAE and DST (India);
Colciencias (Colombia);
CONACyT (Mexico);
KRF and KOSEF (Korea);
CONICET and UBACyT (Argentina);
FOM (The Netherlands);
STFC and the Royal Society (United Kingdom);
MSMT and GACR (Czech Republic);
CRC Program and NSERC (Canada);
BMBF and DFG (Germany);
SFI (Ireland);
The Swedish Research Council (Sweden);
and
CAS and CNSF (China).
%


\begin{thebibliography}{99}
\bibitem{PDL} The event-by-event value of the proper decay length is defined as $L_{xy}M/p_{T}$, where $p_T$ and $M$ are the transverse 
momentum  and mass of the $b$ hadron, respectively, and  $L_{xy}$ is the  distance between the $p\bar{p}$ interaction vertex 
and the $b$ hadron decay vertex in the transverse plane.
\end{thebibliography}
\end{document}